\documentclass[reprint, amsmath, amssymb, aip, cha]{revtex4-1}
\usepackage{graphicx}
\usepackage{dcolumn}
\usepackage{bm}

\begin{document}
\title{Noise induced stabilization of chaotic free-running laser diode}

\author{Martin Virte}
\email{Corresponding author: mvirte@b-phot.org}
\affiliation{Brussels Photonics Team, Dept. of Applied Physics and Photonics, Vrije Universiteit Brussel, Pleinlaan 2, 1050 Brussel, Belgium}

\date{\today}


\begin{abstract}
In this paper, we investigate theoretically the stabilization of a free-running vertical-cavity surface-emitting laser exhibiting polarization chaos dynamics. We report the existence of a boundary isolating the chaotic attractor on one side and a steady-state on the other side, and identify the unstable periodic orbit playing the role of separatrix. In addition, we highlight a small range of parameters where the chaotic attractor passes through this boundary, and therefore where chaos only appears as a transient behaviour. Then, including the effect of spontaneous emission noise in the laser, we demonstrate that, for realistic levels of noise, the system is systematically pushed over the separating solution. As a result, we show that the chaotic dynamics cannot be sustained unless the steady-state on the other side of the separatrix becomes unstable. Finally, we link the stability of this steady-state to a small value of the birefringence in the laser cavity and discuss the significance of this result on future experimental work.
\end{abstract}


\maketitle

\textbf{Semiconductor lasers - or laser diodes - are small, efficient and cheap laser devices and therefore are widely used for industrial applications. To generate a chaotic output however, since they typically behave as damped oscillators, an external perturbation such as optical feedback or modulation is required\cite{Sciamanna2015}. But recently, it has been shown that some specific semiconductor laser structures - namely Vertical-Cavity Surface-Emitting lasers (VCSELs) - could generate chaotic polarization fluctuations without the need for an external forcing due to a competition between two modes in the laser cavity\cite{Virte2012}.\\
The specific dynamics of VCSELs has been studied intensively for more than 20 years\cite{Chang-hasnain1991, Choquette1994, SanMiguel1995, Martin-Regalado1997, Panajotov1998, Willemsen1999, Ryvkin1999, Nagler2003, Panajotov2013}, and polarization chaos has only been observed once, recently and only in nanostructured devices. Yet the theoretical framework reproducing accurately the observed dynamics does not take the specificities of the nanostructures into account\cite{SanMiguel1995, Martin-Regalado1997}, which therefore raises the question: why has this dynamics never been observed in standard, commercial VCSELs before?\\
Here we provide some elements of answer to this question. We theoretically highlight the existence of a separatrix between the chaotic dynamics and a steady-state, and demonstrate that the noise easily pushes the system over this boundary which therefore suppresses the chaotic dynamics. Moreover, we show that this mechanism appears for a large range of parameters, and in particular when the birefringence of the laser cavity is small. Finally, we discuss the relevance of this result by comparing available experimental data between chaotic and stable devices, and show that the chaos suppression mechanism described here is coherent with experimental reports.}

\section{Introduction}
Besides their well-known advantages such as a low threshold and a high bandwidth, vertical-cavity surface-emitting lasers (VCSELs) also attracted interest in the last twenty years due to their polarization instabilities \cite{Chang-hasnain1991, Choquette1994}. Although they generally emit linearly polarized (LP) light at threshold, an increase of the injection current or a change in temperature can induce a switching to the orthogonal linear polarization, a so-called polarization switching, see e.g. \cite{Panajotov2013} for a recent review. These polarization changes were typically interpreted as thermal or noise-induced processes \cite{Panajotov1998, Willemsen1999, Ryvkin1999, Nagler2003} until a chaotic behaviour has been unambiguously identified as the result of the polarization mode competition \cite{Virte2012}. Although this last result confirmed the long-standing prediction of the dynamical spin-flip model (SFM) \cite{SanMiguel1995, Martin-Regalado1997}, the experimental conditions required to obtain such peculiar dynamics are still unclear. So far, polarization chaos has only been observed in the high-speed quantum dot VCSELs described in \cite{Hopfer2006}. Even though some earlier observations in quantum well VCSELs might be consistent with the chaotic dynamics observed \cite{Ackemann2001}, there is to the best of our knowledge no clear report of a similar behaviour in commercial devices. On the other hand, the spin-flip model does not integrate any specific mechanism related to quantum dots as gain medium but still provides an accurate qualitative description of the chaotic dynamics. Thus it suggests that the quantum dot nature of the gain medium might not be the key element leading to chaos. Further investigations are therefore required to better understand the origin of the chaos and of the factors limiting its occurrence in commercial quantum well devices.\\
Here, we theoretically report and investigate a new mechanism destroying the chaotic dynamics and which dramatically reduces the range of parameters for which polarization chaos can be observed when realistic levels of spontaneous emission noise are considered. This contribution is organized as follows: first we present in detail the theoretical model derived from the SFM and its parameters. Second, using direct numerical integration, we investigate three different cases exhibiting significantly different dynamical scenarios, and highlight the disappearance of the chaos for a certain range of parameter. Third, using continuation techniques, we bring further details about the three selected cases and clarify the mechanism leading to the suppression of the chaotic dynamics. Fourth, we introduce noise terms in our modelling to take into account the effect of the spontaneous emission and show that low levels of noise are sufficient to destroy the chaotic dynamics via the same mechanism but in a much larger range of parameters. In the fifth section, based on the theoretical insight of the previous sections, we discuss the impact that such mechanism could have experimentally and confront it with the data of previous experimental observation of polarization chaos and polarization switching events. Finally, in the last part, we conclude and summarize our results and their potential interest for further investigations.

\section{Theoretical model SFM}
The work described in this letter is based on the spin-flip model (SFM) for VCSELs \cite{SanMiguel1995, Martin-Regalado1997} which predicts accurately the emergence of the polarization chaos dynamics \cite{Virte2012}. Here however we use a slightly more complex version of this model - introduced in \cite{Travagnin1997} - in which misaligned phase and amplitude anisotropies are considered. In the Stokes parameter phase space, the polarization chaos forms a double-scroll attractor with its two scrolls rotating around two elliptically polarized steady-state. In the original SFM framework, these two scrolls and two steady-states - along with all the other intermediate states that appears on the route to chaos - are perfectly symmetrical. But this is not completely accurate in practice as asymmetries can easily be observed experimentally \cite{Virte2014, Virte2015}. Including an anisotropy misalignment therefore breaks the symmetry of the system, which allows to reproduce detailed features of the dynamical scenario and to ensure a higher level of generality. So the model considered here reads as: 
\begin{align}
\frac{dE_\pm}{dt} = & \kappa (1+i\alpha)(N\pm n-1) E_\pm \nonumber \\
& - (i\gamma_p+(cos(2\theta) \mp i sin(2\theta))\gamma_a)E_\mp \label{eq:SFMfield} \\
\frac{dN}{dt} = &-\gamma \left( N -\mu + (N+n)|E_+|^2 + (N-n)|E_-|^2 \right) \\
\frac{dn}{dt} = &-\gamma_s n -\gamma \left( (N+n)|E_+|^2 - (N-n)|E_-|^2 \right)
\end{align}
with $E_{\pm}$ the left (-) and right (+) circular polarizations, $N$ the total carrier population and $n$ the carrier population difference between the two carrier reservoirs considered for the two separated emission processes \cite{Martin-Regalado1997}. The different time-scales are modelled by the decay rate of the electric field $\kappa$, the carrier decay rate $\gamma$ and the spin-flip relaxation rate $\gamma_s$. The anisotropies inside the laser cavity are defined by the phase anisotropy (or birefringence) $\gamma_p$, the amplitude anisotropy $\gamma_a$ and the misalignment angle $\theta$ between the axis of maximum frequency and the axis of maximum losses. $\alpha$ is the linewidth enhancement factor and the injection current is represented by $\mu$. 
Unless specified otherwise, we use the following parameter values which are similar to those used in previous works \cite{Virte2013, Virte2014, Virte2015}: $\alpha = 3$, $\kappa = 600 \, ns^{-1}$, $\gamma = 1  \, ns^{-1}$, $\gamma_s = 100 \, ns^{-1}$, $\gamma_a = -0.7  \, ns^{-1}$, $\theta = 0.05 \, rad$. Although the chaotic dynamics appears using these values, it is important to emphasize that chaos is obtained in a large range of parameters and not only in a small region of the parameter space. In particular, a misalignment between the phase and amplitude anisotropy is not required to observe polarization chaos as already highlighted in previous reports \cite{Virte2012, Virte2013}.
When dealing with polarization, it is often convenient to use Stokes parameters that we designate as $s_0$, $s_1$, $s_2$ and $s_3$, and which can be normalized as: $s_1/s_0$, $s_2/s_0$ and $s_3/s_0$ \cite{Martin-Regalado1997}. With this normalization, the system trajectory lies on a sphere of radius 1 in a three dimensional phase-space.

\begin{figure}
\includegraphics[width = \linewidth]{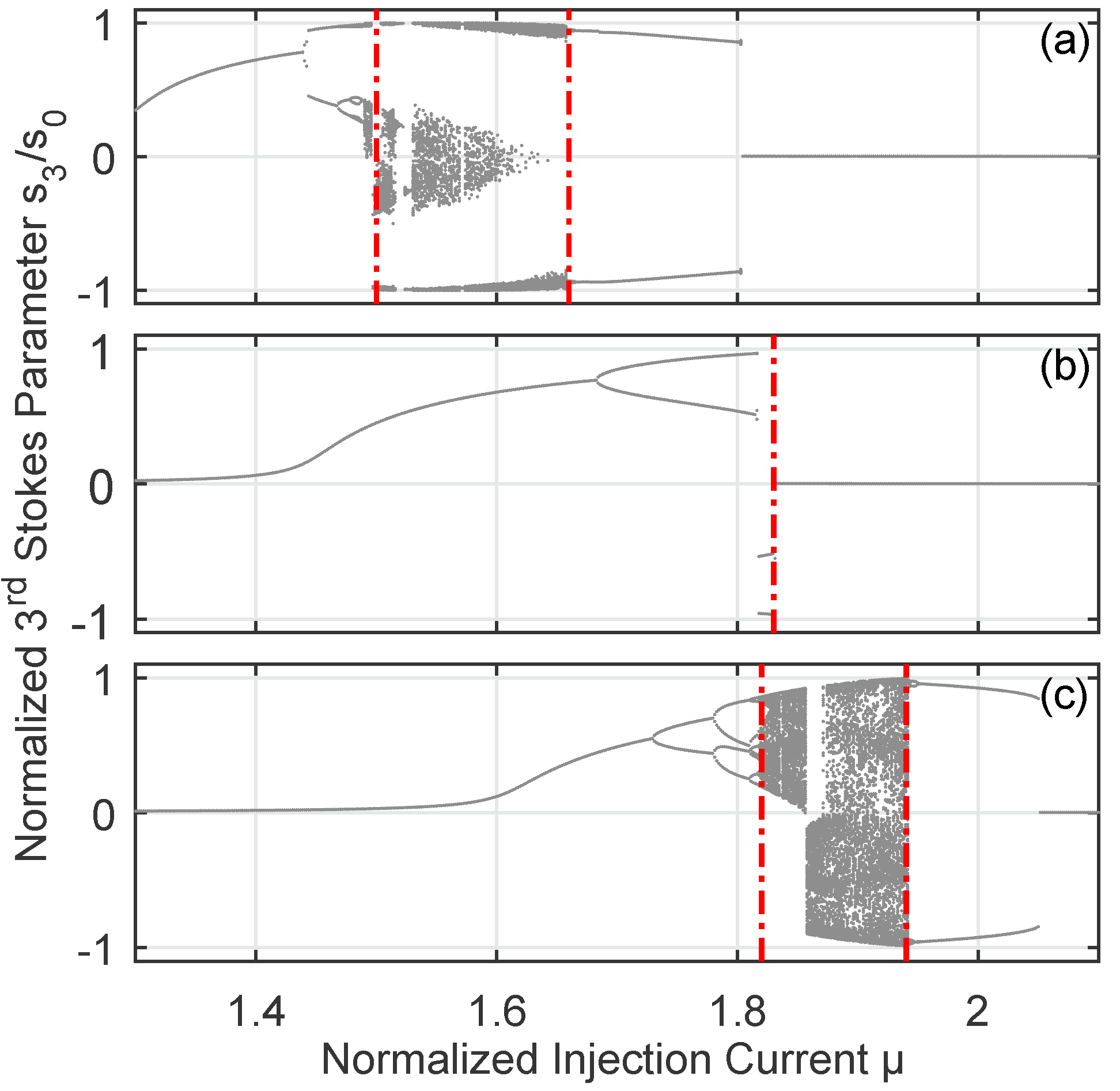}
\caption{Bifurcation diagram showing the extrema of the normalized $3^{rd}$ Stokes parameter ($s_3/s_0$) time-series for increasing injection currents and for three different birefringence values (a) $\gamma_p = 5 \, ns^{-1}$, (b) $\gamma_p = 8 \, ns^{-1}$ and (c) $\gamma_p = 11 \, ns^{-1}$. The vertical red dash-dotted lines highlight the limits of the chaotic regions, except in (b) where it indicates the switching point toward the YLP steady-state. \label{fig:SimRes1}}
\end{figure}

\section{Evolution of the chaotic dynamics}
In Fig. \ref{fig:SimRes1}, we show bifurcation diagrams depicting the evolution of the VCSEL dynamics for increasing injection currents. For each of the three cases we focus on the normalized third Stokes parameter $s_3/s_0$. The main interest of this projection is that it clearly separates the two scrolls of the polarization chaos attractor as one shows $s_3 > 0$ and the other $s_3 < 0$. Thus, plotting the extrema of its time-series, we can discriminate the two elliptical polarization orientation of the system and observe the evolution from steady-states - where a single point is obtained - up to complex dynamical states such as quasi-periodic dynamics and chaos - for which numerous points are recorded.\\
In all cases highlighted here, the laser emits linearly polarized light at threshold ($s_3 = 0$) - defined as the XLP state - before experiencing a transition through elliptical polarization. Due to the anisotropy misalignment with an angle of $\theta = 0.05$, the transition toward the elliptically polarized state is smooth and we quickly reach $s_3 > 0$. When $\theta$ is negative, the orientation is simply reversed with $s_3 < 0$. Then, the laser is destabilized toward dynamical behaviour until it reaches the YLP state, i.e. the linear polarization orthogonal to the XLP state stable at threshold, for which we again have $s_3 = 0$. 
Within the dynamical region, the influence of the anisotropy misalignement is quite clear as it induces switching(s) between the two elliptical polarization corresponding to $s_3>0$ and $s_3<0$ respectively \cite{Virte2014}.
As shown in Fig. \ref{fig:SimRes1}, the laser always experiences a dynamical transition before reaching the YLP steady-state for larger injection currents, but we can clearly observe a severe qualitative change as the birefringence is increased from (a) $\gamma_p = 5 \, ns^{-1}$, to (c) $\gamma_p = 11 \, ns^{-1}$. First obvious observation is that the chaotic dynamics is completely absent in case (b). Indeed, for $\gamma_p = 8 \, ns^{-1}$, the system only exhibit simple oscillations - although with a switching of the polarization orientation - before it reaches the final YLP stationary state. Secondly, we can also see that the chaotic dynamics itself is significantly altered. In (a), we observe numerous points gathered around three distinct values $(-1, \, 0, \, 1)$, which suggest a well-defined chaotic attractor with a eight-shape similar to the one initially identified for the polarization chaos dynamics \cite{Virte2012}. On the other hand, in Fig. \ref{fig:SimRes1}(c), this observation does not hold anymore as the points are scattered in the full range of possible values, from -1 to 1. And indeed, with a projection in the Stokes parameter space, we can confirm some qualitative changes of the chaotic attractor with the two scrolls looking now more like two vortices connected around $S_3 \sim 0$. Although the qualitative differences of the two bifurcation scenarios have been described in \cite{Virte2013}, almost no details regarding the impact on the chaotic dynamics has been presented in this report. Further details are however out of the scope of this study.\\

\begin{figure}
\includegraphics[width = \linewidth]{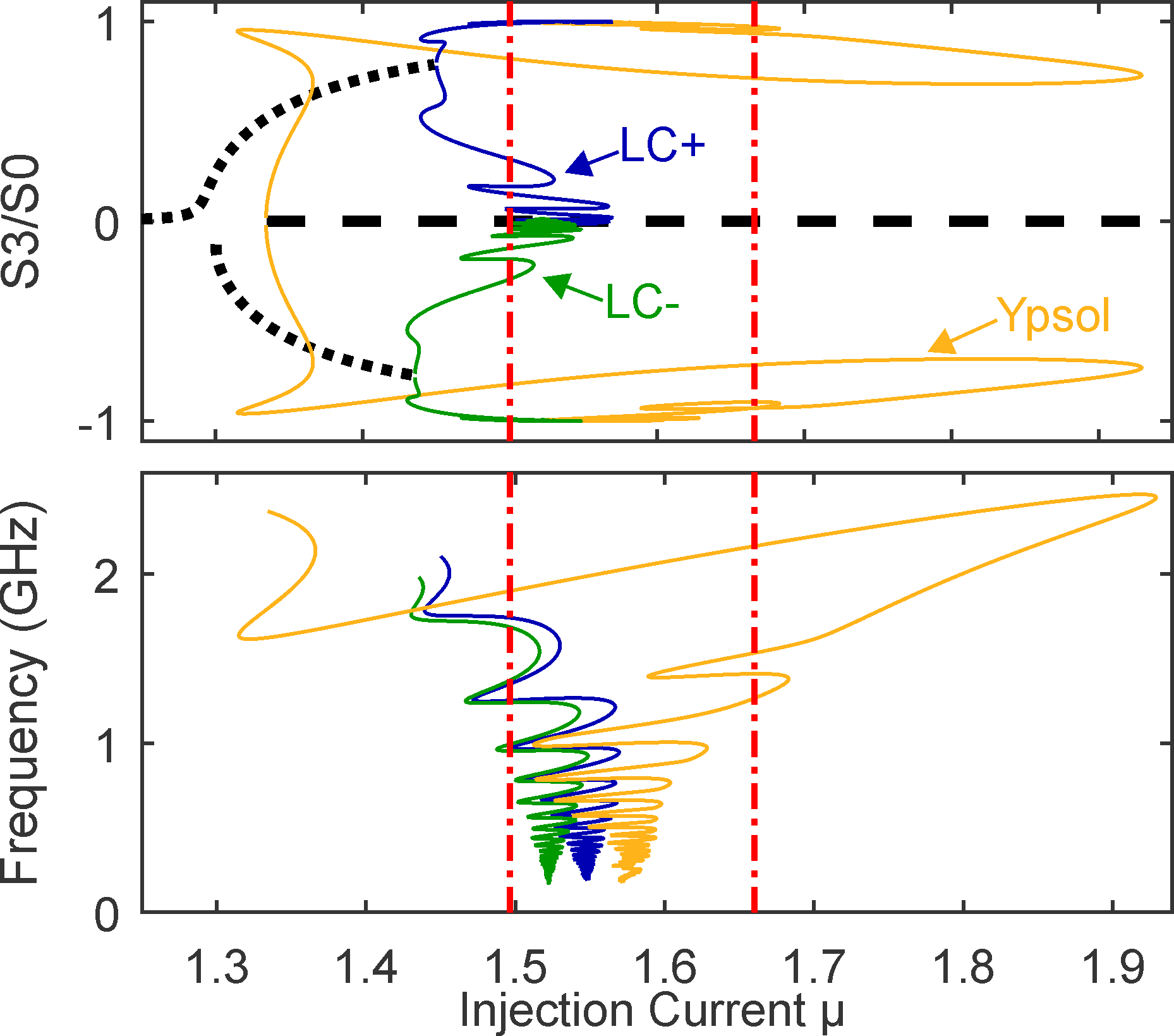}
\caption{Bifurcation diagram showing the solutions of the system projected on the $3^{rd}$ Stokes parameter (top) and their frequency (bottom). Only stable steady-states are shown: elliptically polarized states (dotted) and Y-LP (dashed). The three other branches are periodic solutions ($LC+$, $LC-$ and $Ypsol$, see text for details) for which we plot the maximum and minimum value of their $s_3/s_0$ projection (top) along with their frequencies (bottom). Vertical red dash-dotted lines show the limits of the chaotic region obtained with direct numerical integration shownin Fig. \ref{fig:SimRes1}. For clarity, the bifurcations and the stability of branches are not shown. \label{fig:ContCase1}}
\end{figure}

\begin{figure}
\includegraphics[width = \linewidth]{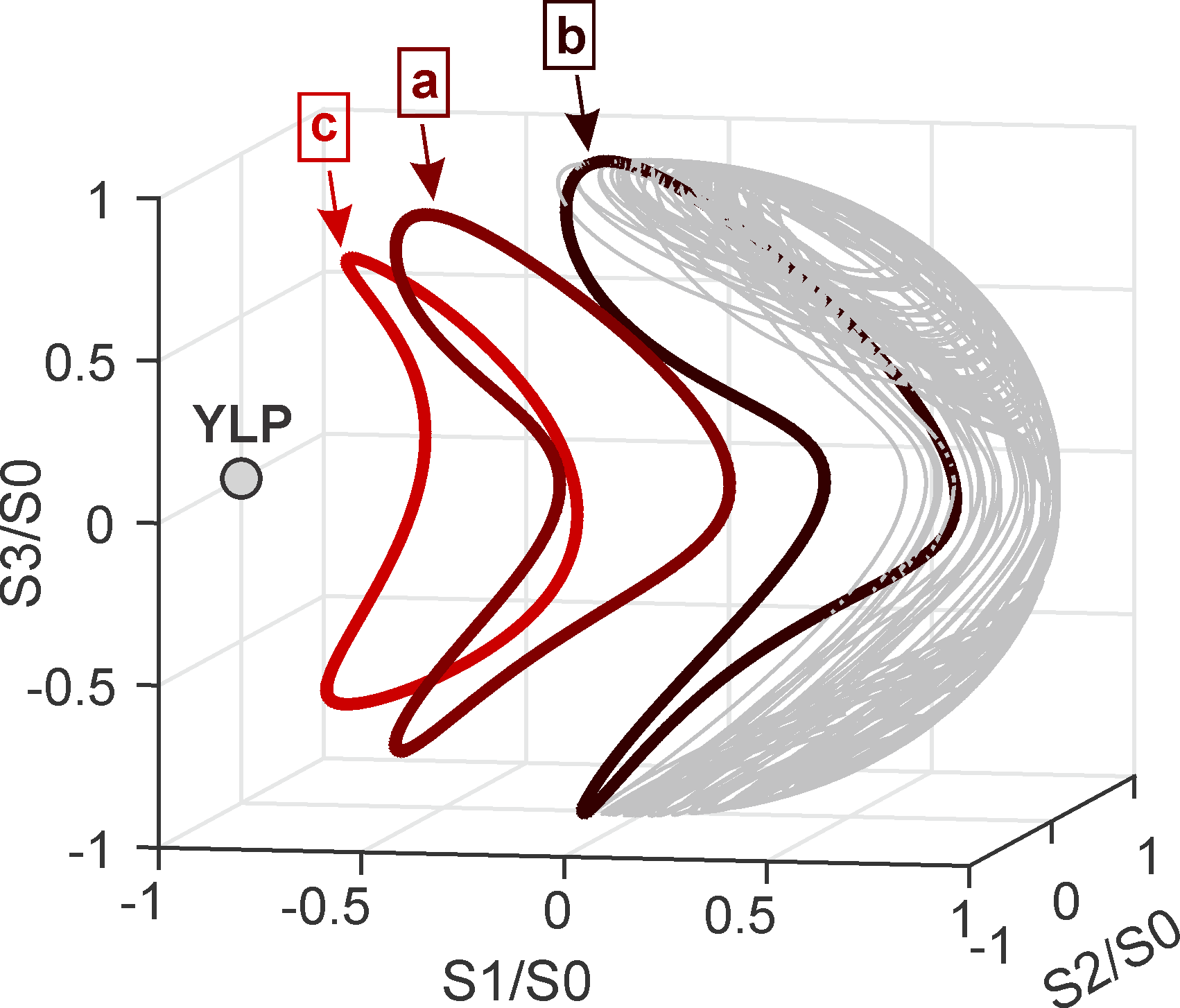}
\caption{Representation of the barrier orbits in the normalized Stokes parameter phase space. The point on the left side correspond to the YLP steady-state while a representation of the chaotic attractor - obtained for $\mu = 1.74$ and $\gamma_p = 7.2 \, ns^{-1}$ - is shown on the right side. In between, we plot three orbits representative of each case - as shown by the labels - obtained for: (a) $\gamma_p = 5 \, ns^{-1}$ and $\mu = 1.5$, (b) $\gamma_p = 8 \, ns^{-1}$ and $\mu = 1.8$, (c) $\gamma_p = 11 \, ns^{-1}$ and $\mu = 2$. \label{fig:3Dplot}}
\end{figure}

To understand the origin of the chaos disappearance, we take advantage of continuation techniques to follow both stable and unstable solutions. Here, we use the dde-biftool toolbox \cite{Engelborghs2001, Sieber2015}. Using this approach, we can have a complete picture of existent stable and unstable solutions and of the different elements required to obtain the polarization chaos dynamics as shown in Fig. \ref{fig:ContCase1}. In addition to the steady elliptically polarized states and the YLP solution, we were able to follow three periodic solutions created by Hopf bifurcations on the three steady-states. We identify these solutions as $LC+$ ($LC-$) for the orbit oscillating around the elliptically polarized solution with positive (negative) value of $s_3$, and $Ypsol$ for the periodic solution born from the Hopf bifurcation on the YLP steady-state branch.\\
From the strong decrease of their frequency tending to zero, we can conclude that all three periodic solutions collide with a steady-state and form so-called Shilnikov bifurcations. Interestingly, all branches are colliding with the unstable linear polarization XLP at $s_3 = 0$, i.e. the linear polarization stable at threshold. This collision is not clear for the Ypsol branch based on Fig. \ref{fig:ContCase1} because its projection on the first normalized Stokes parameter cover almost the whole range of accessible values $[-1, \; 1]$ and only the extrema are shown, but an observation of the complete orbit in the three dimensional Stokes parameter clearly shows this feature. Then, we also observe - as already reported in \cite{Virte2013} - that the orthogonal linearly polarized state YLP is stable while the system exhibit chaotic dynamics. From this observation we can easily deduce that the chaotic attractor can only exist 
if a separatrix exists and isolates the chaotic dynamics from the basin of attraction of the steady-state. By projecting available steady-states and periodic solutions in the normalized Stokes parameter phase space, we can easily identify the unstable periodic solution Ypsol as the one playing the role of the sepratrix. In Fig. \ref{fig:3Dplot}, we show a projection of this periodic orbit - identified as orbit (a) - along with the projection of a typical chaotic attractor trajectory and the YLP steady-state.

\begin{figure}
\includegraphics[width = \linewidth]{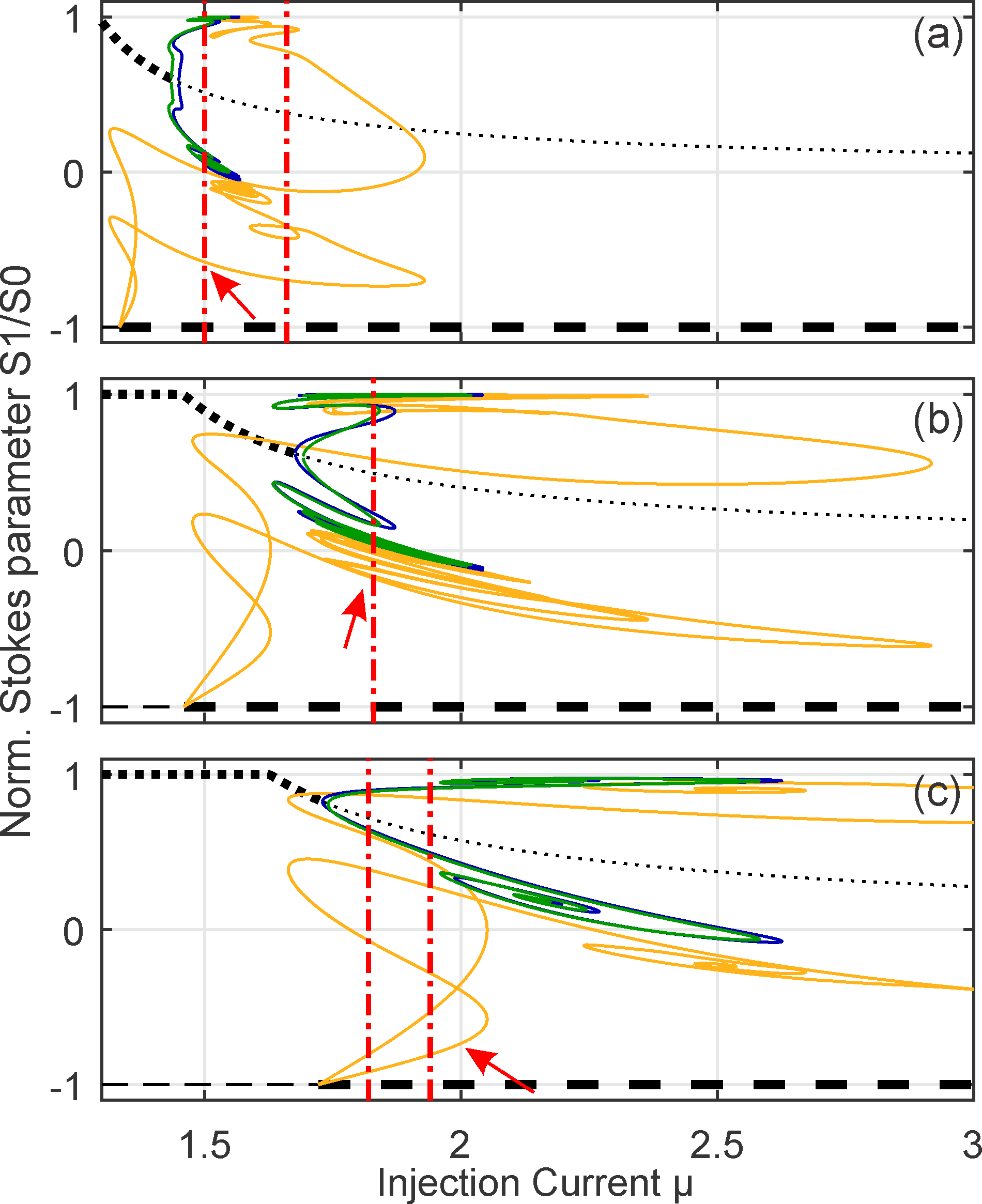}
\caption{Bifurcation diagram of the system solutions projected on the $1^{st}$ normalized Stokes parameter versus injection current for the three cases discussed in Fig. \ref{fig:SimRes1}. The solutions are represented as described in the caption of Fig. \ref{fig:ContCase1}. The vertical red dash-dotted lines indicate the limit of the chaotic region, except for (b) where it shows when the laser switches to the Y-LP steady-state. For each panel, the arrow shows the current value for which the periodic orbit shown in Fig. \ref{fig:3Dplot} is obtained. \label{fig:S1plot}}
\end{figure}

\section{Chaos breaking mechanism}
When we apply continuation techniques to the different cases highlighted in Fig. \ref{fig:SimRes1}, we surprisingly observe that all the elements identified previously are still present, see Fig. \ref{fig:S1plot}. The three periodic solutions emerge in a similar manner, and are also destroyed through Shilnikov bifurcations. In addition, no significant difference in terms of stability or eigenvalues of the periodic orbits is observed. However, when we compare direct numerical integration and continuation results as the birefringence is increased we can observe that the chaotic dynamics only reappears once the lower part of the Ypsol branch - identified by the arrow in Fig. \ref{fig:S1plot}(c) - goes beyond the point when the systems switches to the YLP steady state: i.e. the point beyond which complex dynamics are expected. Such observation therefore indicates that this part of the Ypsol branch has become essential to obtain chaos while it was not a requirement before. In fact, this new orbit now plays the role of the separatrix between the chaotic dynamics and the stable YLP steady-state, as discussed below.\\
In Fig. \ref{fig:3Dplot}, we clearly see that the chaotic trajectory is approximately restricted to positive values of $s_1$, while the stable YLP steady-state is located at the opposite side at $s_1/s_0 = -1$. Obviously, the barrier orbit needs to be in between in order to play its role effectively. However, when we look at the evolution of the orbit position projected on $s_1$, as in Fig. \ref{fig:S1plot}, we clearly see that the barrier orbit in (a) moves toward higher values of $s_1$ in (b) and (c), i.e. closer and closer to the chaotic dynamics. This is even clearer in the Stokes parameter space as shown in Fig. \ref{fig:3Dplot}. Between case (a) and (b), the barrier orbit dramatically moves toward the chaotic attractor. Hence, as a result of this close proximity, we see that the chaotic trajectory goes beyond the barrier orbit: in this situation the chaotic dynamics will only appear as a transient behaviour before settling to the YLP steady-state.\\
As already mentioned, for larger birefringence values - case (c) - the lower part of the Ypsol branch becomes available in the dynamical region. In the Stokes parameter projection of Fig. \ref{fig:3Dplot}, we see that the new orbit is ideally located to isolate the chaotic dynamics, and thus replaces the other orbit as a separatrix. In addition, we also see in Fig. \ref{fig:S1plot} that the chaotic dynamics emerges far from the three Shilnikov bifurcations in comparison to case (a). In particular, it means that the distance between the two unstable limit cycles $LC+$ and $LC-$ supporting polarization chaos is much larger than in case (a), and therefore that the jumps between the two polarization orientations will not be as smooth. We can also expect the dynamics to be influenced by the very close Ypsol orbit, i.e. the one formerly separating the two dynamics for lower values of the birefringence. All these discrepancies are consistent with the qualitative differences that can be observed between the polarization chaos dynamics in case (a) and (c) which has already been briefly discussed in the previous section.\\
For even larger values of the birefringence, the Y-LP steady-state becomes stable only for much larger - and most often unrealistic - injection currents \cite{Virte2013} which thus remove the bistability between the YLP steady-state and the chaotic dynamics. Thus the need for a barrier solution is removed and the chaos is systematically observed.

\section{Impact of the spontaneous emission noise}
So far, we analysed theoretically the spin-flip model for VCSELs without considering any noise contribution while in practise noise - and in particular spontaneous emission noise - is not only unavoidable, but also a relatively strong driving force in semiconductor lasers. In the context of this contribution, it is clear that a sufficiently strong noise will be able to kick the system out of the chaotic attractor over the separating orbit, hence will destroy the chaotic dynamics and leave the system on the YLP steady-state. It is therefore crucial to evaluate whether the scenarios described previously can withstand realistic noise levels - and might potentially be observed experimentally - or if the noise becomes the dominant driving force.
\\

\begin{figure}
\includegraphics[width = \linewidth]{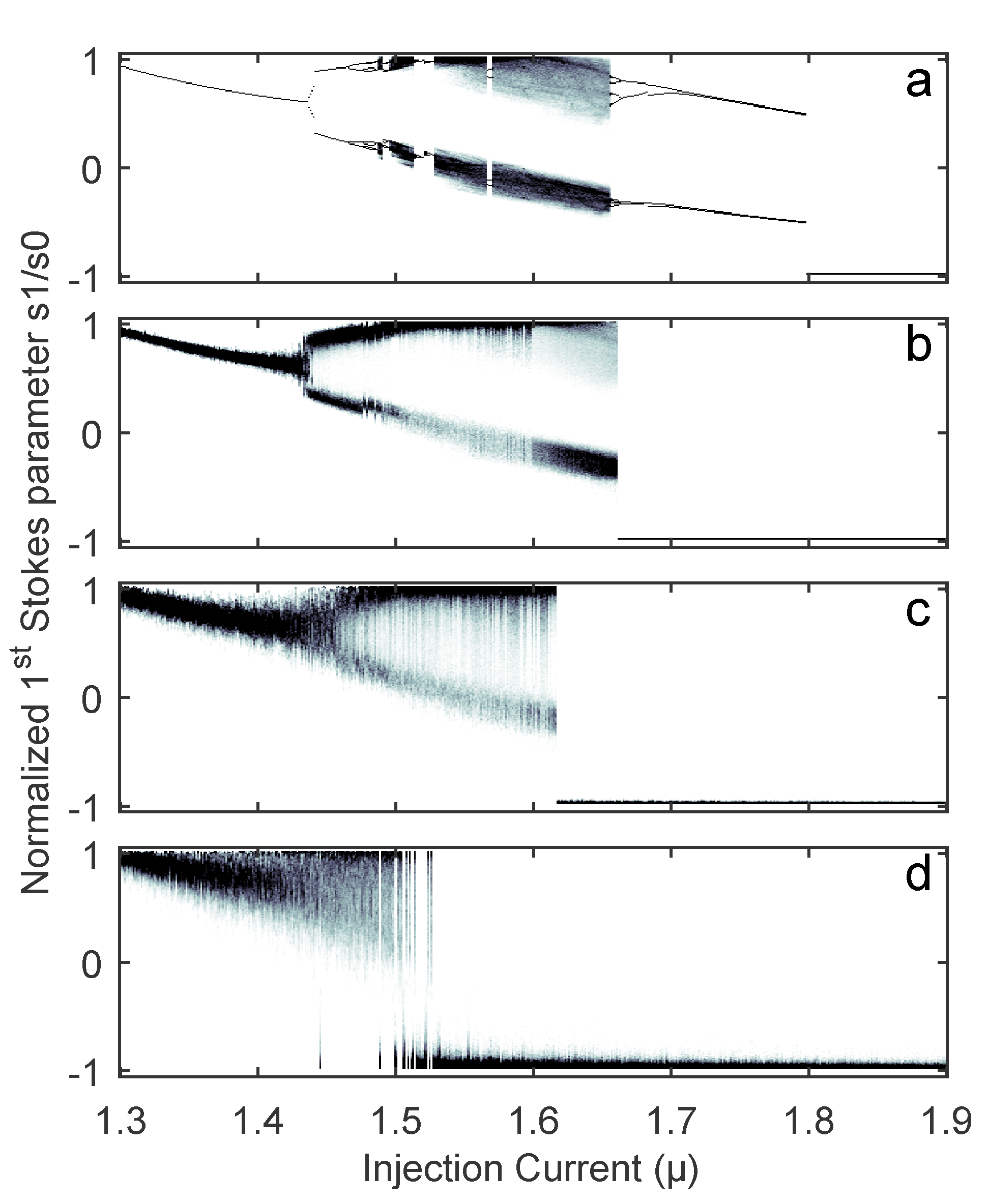}
\caption{Evolution of the histogram of simulated noisy time-series extrema for increasing injection current and for $\gamma_p = 5 \, ns^{-1}$ and spontaneous emission rates of $\overline{\beta_{sp}} = 10^{-15}\, ns^{-1}$ (a), $10^{-7}\, ns^{-1}$ (b), $10^{-6}\, ns^{-1}$ (c) and $10^{-5}\, ns^{-1}$ (d) respectively. All other parameters are similar to those used in the previous sections.\label{fig:noise_gamp5}}
\end{figure}

For this analysis, we introduce an additional term in eq. \ref{eq:SFMfield} corresponding to the spontaneous emission noise in the laser cavity; a detailed derivation of this term can be found e.g. in section 3.5 of Ref. \cite{Ohtsubo2012}. The field equations then become:
\begin{align}
\frac{dE_\pm}{dt} = &\kappa (1+i\alpha)(N\pm n-1) E_\pm \nonumber\\ 
 & - (i\gamma_p+(cos(2\theta) \mp i sin(2\theta))\gamma_a)E_\mp \nonumber \\
 & + \sqrt{\overline{\beta_{sp}}\gamma(N \pm n)}\xi_{\pm}
\end{align}
with $\xi_{\pm}$ two uncorrelated complex white noises with zero mean and unitary variance. $\overline{\beta_{sp}}$ is the spontaneous emission rate expressed in $ns^{-1}$ and defined as $\overline{\beta_{sp}} = \beta / 2\Delta t$; $\beta$ is the spontaneous emission factor corresponding to the fraction of spontaneous emission noise that enters the lasing mode \cite{Coldren2012}, and $\Delta t$ the simulation time-step. In this contribution, we use a simulation time-step of $\Delta t = 1 \, ps$. Typical values for the spontaneous emission factor $\beta$ are about $10^{-5}$ for edge-emitting lasers and $10^{-4}$ for VCSELs mostly due to their reduced cavity length \cite{Coldren2012}. With a simulation time-step of $1\, ps$, these values correspond to $\overline{\beta_{sp}} = 5.10^{-3} \, ns^{-1}$ and $5.10^{-2} \, ns^{-1}$ respectively. Adding a noise contribution in our modelling also impacts our analysis tools: continuation techniques cannot be used and bifurcation diagrams become inaccurate as additional extrema appear in the time-series due to the noise contribution. To overcome this hurdle, instead of considering directly the extrema of the time-series, we focus on the histogram of these extrema. In practice that means we need to acquire a much larger set of values for the extrema and therefore that we must simulate much longer time-series.\\

\begin{figure}
\includegraphics[width = \linewidth]{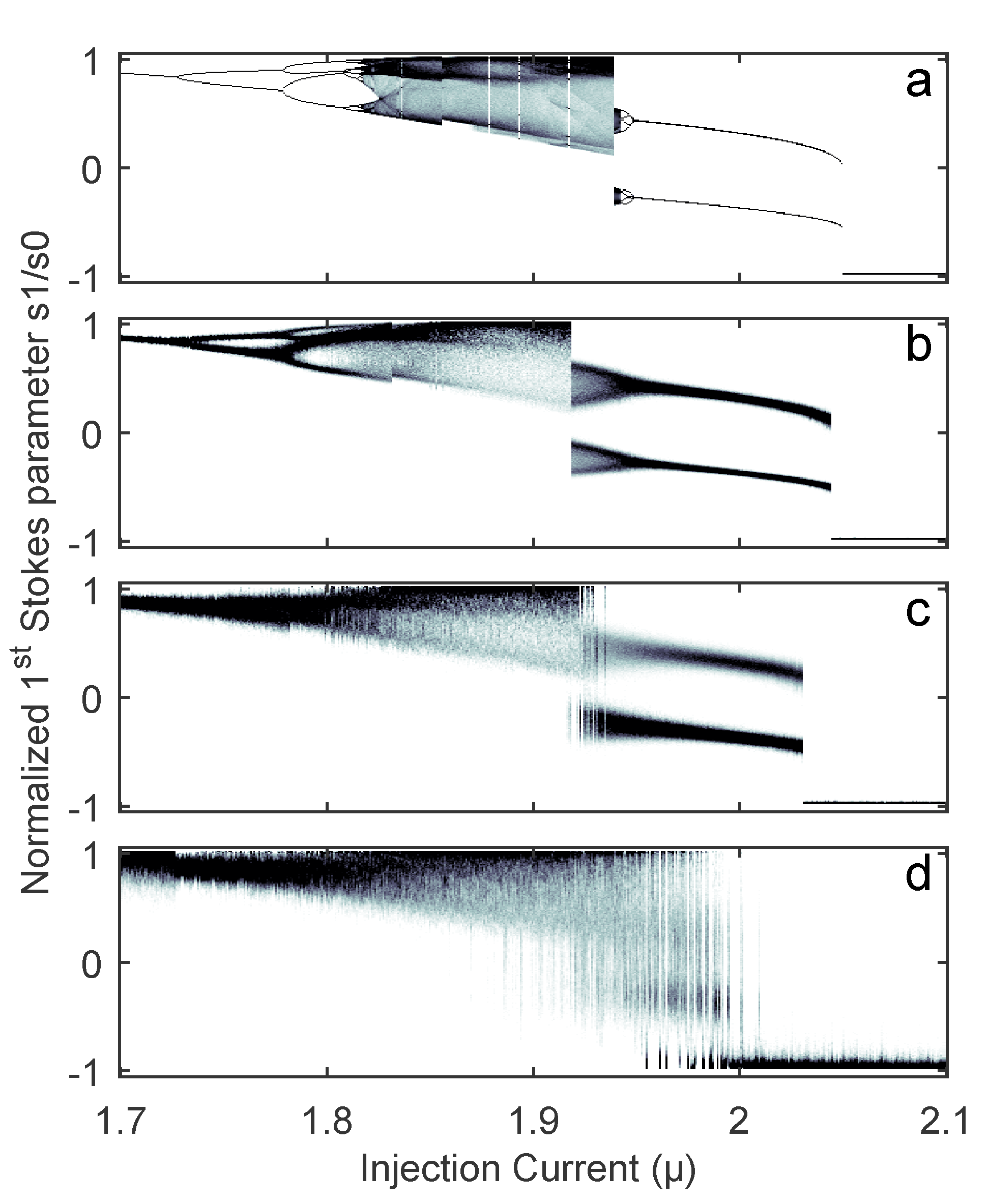}
\caption{Evolution of the histogram of simulated noisy time-series extrema for increasing injection current and for $\gamma_p = 11 \, ns^{-1}$ and spontaneous emission rates of $\overline{\beta_{sp}} = 10^{-15}\, ns^{-1}$ (a), $10^{-7}\, ns^{-1}$ (b), $10^{-6}\, ns^{-1}$ (c) and $10^{-5}\, ns^{-1}$ (d) respectively. All other parameters are similar to those used in the previous sections. \label{fig:noise_gamp11}}
\end{figure}
With this approach we obtain the results displayed in Fig. \ref{fig:noise_gamp5}, where we show the evolution of the dynamics for a spontaneous emission rate of $\overline{\beta_{sp}} = 10^{-15}\, ns^{-1}$, $10^{-7}\, ns^{-1}$, $10^{-6}\, ns^{-1}$ and $10^{-5}\, ns^{-1}$ in panels (a) to (d) respectively. It is worth noting that these values correspond to a spontaneous emission factor $\beta$ at least three to five orders of magnitude below the typical values expected in a real structure. Unlike Fig. \ref{fig:SimRes1}, we consider the first Stokes parameter versus the injection current in order to clearly see when the system switches from chaotic behaviour - i.e. $s_1/s_0 > -0.5$ in our case - to the YLP steady-state $s_1/s_0 = -1$. In (a), the noise level is extremely small and is therefore taken as a ``no-noise'' reference point. We observe that in this case all the dynamical details highlighted in the previous sections are accounted for and no impact of the noise can be spotted at this point. When the spontaneous emission rate is increased however, we see that the bifurcation diagram becomes blurry, see Fig. \ref{fig:noise_gamp5}(b-d): steady-state branches are getting wider and the most precise details disappear. For $\overline{\beta_{sp}} = 10^{-7}\, ns^{-1}$ - displayed in Fig. \ref{fig:noise_gamp5}(b) - we already see some significant changes in the evolution of the laser dynamics. For currents above $\mu \approx 1.65$, the laser directly settles on the YLP steady-state instead of exhibiting periodic oscillations up to $\mu \sim 1.8$. The chaotic region is maintained, but for larger spontaneous emission rate, this region gradually shrinks until it vanishes completely. Thus, for $\overline{\beta_{sp}} = 10^{-5}\, ns^{-1}$ polarization chaos does not appear at all and the laser just experiences a type II polarization switching with a transition through elliptically polarized state and P1 oscillations.\\

We showed in the previous sections that for larger values of the birefringence, the periodic orbit separating the two dynamics is much closer to the Y-LP steady-state, and therefore further away from the chaotic attractor. Hence we can expect that for  $\gamma_p = 11 \, ns ^{-1}$, polarization chaos will be more robust against the noise perturbation. Simulation results tend to confirm this point of view as can be seen in Fig. \ref{fig:noise_gamp11}. Indeed, with this larger birefringence, we see that the influence of the noise is strongly reduced in comparison to the previous case. For the lower level of spontaneous emission rate considered - $\overline{\beta_{sp}} = 10^{-7}\, ns^{-1}$ and $\overline{\beta_{sp}} = 10^{-6}\, ns^{-1}$ in Fig. \ref{fig:noise_gamp11}(b) and (c) respectively - we mostly observe the appearance of a blur of the bifurcation diagram but all the elements of the dynamical scenario remain present. In particular, the periodic solution after the chaotic region is clearly more robust against the noise perturbation. However for $\overline{\beta_{sp}} = 10^{-5}\, ns^{-1}$, in Fig. \ref{fig:noise_gamp11}(d), this periodic solution start to be severely impacted but still exist. We also see that the transition from chaos to P1 dynamics becomes unclear, but the chaotic behaviour is conserved even though its detailed features are wiped off by the noise. Nevertheless, although a separating orbit closer to the steady-state solution seem to improve the robustness of the system against the noise, the high level of spontaneous emission noise typically expected in VCSEL devices will likely erase most dynamical features.

\section{Comparison with experimental observations \label{sec:compXP}}
\begin{figure}
\includegraphics[width = \linewidth]{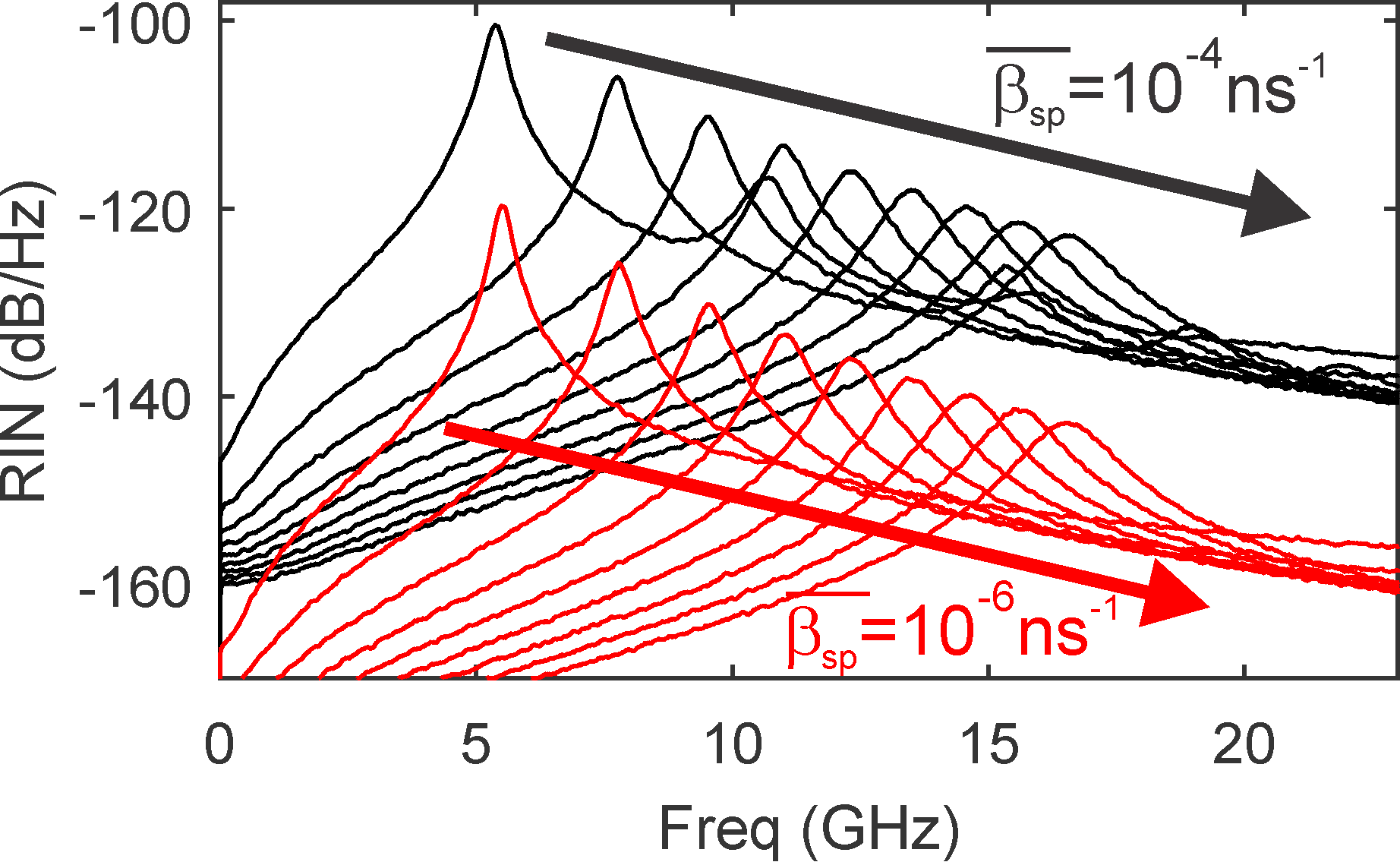}
\caption{Evolution of the RIN for increasing injection currents from $\mu = 2$ up to $10$ (from left to right, following the arrow), and for two spontaneous emission rate of $\overline{\beta_{sp}} = 10^{-4}\, ns^{-1}$ (top, black) and $\overline{\beta_{sp}} = 10^{-6}\, ns^{-1}$ (bottom, red). The results that are displayed here has been computed with a phase anisotropy of $\gamma_p = 5 \, ns^{-1}$, but this parameter seem to have almost no impact on the RIN as the same evolution can be observed for a large range of values. \label{fig:RIN}}
\end{figure}

Based on the previous observations, it seems very unlikely that polarization chaos could be observed experimentally with devices exhibiting a behaviour similar to the one discussed in the previous section. Nonetheless, it is interesting to remark that the emergence of the chaotic dynamics in the cases described in this contribution - in particular the one with the lowest birefringence $\gamma_p =  5 \, ns^{-1}$ - is hindered by two essential features: 1/ the bistability between the chaotic dynamics and the Y-LP steady-state let a stabilization opportunity to the system, and therefore to the disappearance of the chaotic dynamics. As already mentioned in the previous sections, this bistability is removed when birefringence values are sufficiently large e.g. $\gamma_p > 15 \, ns^{-1}$ for the set of parameters considered: in such configuration the Y-LP steady-state only becomes stable after the region of chaotic dynamics. 2/ with the parameters considered in this work, the chaotic dynamics appears for relatively small currents: $\mu < 2$, i.e. relatively close to the laser threshold - here $I_{th} \approx 1$ - which is where the impact of the noise is the strongest. This can typically be quantified by the Relative Intensity Noise (RIN) which is known to be significantly reduced when the injection current is increased \cite{Coldren2012}. In Fig. \ref{fig:RIN}, we display the evolution of the RIN for increasing injection currents with the modified SFM framework used in this contribution. Similar results have been obtained for different values of the birefringence and the anisotropy misalignment; the RIN evolution seems quite independent of these two parameters. Thus we can remark that with a spontaneous emission rate of $\overline{\beta_{sp}} = 10^{-4}\, ns^{-1}$, RIN values as large as $-100 \, dB/Hz$ are obtained at low injection currents. However, for $\mu = 8$ a RIN of $-120 \, dB/Hz$ is achieved which is comparable to the level of noise obtained for $\mu = 2$ and a spontaneous emission rate two order of magnitude lower $\overline{\beta_{sp}} = 10^{-6}\, ns^{-1}$. This could arguably mean that polarization chaos might be observed as in Fig. \ref{fig:noise_gamp5}(c) if the scenario described in the previous sections would take place at injection currents of about 7 or 8 times the threshold.\\
In parallel, if we consider the experimental conditions under which polarization chaos has been observed \cite{Olejniczak2009, Olejniczak2011, Virte2012}, we see that this chaotic dynamics has only been reported at high level of currents $I_{XP} > 7 I_{th}$ and in devices exhibiting a large effective birefringence $\gamma^{eff}_p \approx 25 \, ns^{-1}$, i.e. a large difference of frequency between the two linear and orthogonal preferred polarization modes. Even though the relation between the birefringence in the laser cavity and the measured effective birefringence is not clear, as already briefly discussed in Ref.\cite{Virte2013}, these observations suggest that the experimental conditions were such that the system did not suffer from any of the two problems listed above. On the other hand, previous experimental reports showing some similarities with the scenario leading to chaotic dynamics only considered devices with much smaller birefringence\cite{Ackemann2001, Sondermann2004}, with typical values around $\gamma_p^{eff} < 5 ns^{-1}$, and injection current well below $2I_{th}$. These values are consistent with the parameters considered in this paper, and we can therefore suspect that the chaos suppression mechanism discussed prevented the observation of polarization chaos dynamics in these devices.\\
In the end, considering the bifurcation scenario leading to the emergence of polarization chaos, we know that the region of chaotic dynamics will be pushed toward larger injection currents when the birefringence $\gamma_p$ is increased\cite{Martin-Regalado1997, Virte2013}. Indeed, the starting point of the route to chaos - when no anisotropy misalignment is considered - is the pitchfork bifurcation destabilizing the linear polarization stable at threshold and creating the two elliptically polarized steady-states from which the two scrolls of the chaotic attractor will emerge. In Ref. \cite{Martin-Regalado1997}, the authors demonstrated that this bifurcation takes place at $\mu_x = 1 + \frac{\gamma_s\gamma_p}{\gamma(\kappa\alpha - \gamma_p)}$, which can be approximated as $\mu_x = \frac{\gamma_s}{\gamma\kappa\alpha} \gamma_p$ for typical value of the parameters. Which means that a larger birefringence - with all other parameters unchanged - will lead to a larger value of $\mu_x$ and is therefore expected to both remove the problematic bistability and lead to the chaos arising at larger values of current hence leading to a smaller impact of the noise. Including an anisotropy misalignment destroys the pitchfork bifurcation, but $\mu_x$ still provides a good approximation of the injection current above which dynamical evolution appears. In a nutshell, a larger birefringence in the laser cavity could be expected to remove the two hurdles listed earlier, hence lead to better conditions to generate polarization chaos dynamics. In addition, it should be mentioned that this approach can be implemented in practice by stressing the laser mechanically which allows to tune the birefringence in the cavity to some extent\cite{Pusch2015, Lindemann2016}.

\section{Conclusion}
To conclude, we theoretically report here a new mechanism limiting the range of parameters for which polarization chaos in a free-running VCSEL can be observed. Because the linearly polarized state orthogonal to the polarization at threshold is stable,  polarization chaos can only emerge if one of the unstable periodic orbit of the system isolates the chaotic dynamics from the stable steady-state, i.e. plays the role of a separatrix between the two dynamics. However, we show here that for some values of the birefringence - around $\gamma_p \approx 8 \, ns^{-1}$ with the set of parameters considered - the barrier orbit is too close to the chaotic attractor which therefore goes beyond the limit fixed by this orbit. As a result, the system settles on the stable linearly polarized states after a transient chaotic behaviour. Hence polarization chaos cannot be observed unless another separating orbit appears which happens for larger birefringence values $\gamma_p > 10 \, ns^{-1}$. This re-emergence of chaos is accompanied by a qualitative change in the dynamics: the chaotic attractor becomes more tortuous as the unstable periodic orbits supporting it gets largely separated. When the impact of the spontaneous emission noise is considered, we highlight that for realistic noise levels the separating orbit becomes ineffective and the chaotic dynamics only appears as a transient not only for the problematic cases highlighted for $\gamma_p \approx 8 \, ns^{-1}$, but for the whole range of birefringence value considered $ 5 < \gamma_p < 11$. Based on the experimental data obtained from chaotic devices and previous theoretical investigations, it seems possible to assume that a large birefringence in the laser cavity is requested in order to observe the polarization chaos dynamics in VCSEL devices. Of course, such feature would obviously not be a sufficient characteristic in itself as many other parameters play a significant role in the laser dynamics, but this result could explain at least in part why polarization chaos has not been reported more extensively despite the intensive research on VCSEL dynamics of more than twenty years. This result could also be of practical importance, in particular for the observation of polarization chaos in standard commercial quantum well VCSELs.\\

The author is a Post Doctoral Fellow from the Research Foundation - Flanders (FWO), and acknowledges the support of the METHUSALEM program of the Flemish Government, FWO-Vlaanderen and the inter-university attraction poles program of the Belgian Science Policy Office (IAP P7-35, photonics@be). The author also thanks Krassimir Panajotov, Marc Sciamanna and Hugo Thienpont for fruitful discussions.\\


\end{document}